\documentclass[twocolumn,aps,prb,superscriptaddress,amsmath,amssymb]{revtex4}
\usepackage{graphicx}

\begin{document}

\title{Definitive spectroscopic determination of the transverse interactions responsible for the magnetic quantum tunneling in Mn$_{12}-$acetate}

\author{S. Hill}
\email[corresponding author, Email:]{hill@phys.ufl.edu}
\affiliation{Department of Physics, University of Florida,
Gainesville, FL 32611,USA}
\author{R. S. Edwards}
\affiliation{Department of Physics, University of Florida,
Gainesville, FL 32611,USA}
\author{S. I. Jones}
\affiliation{Department of Physics, University of Florida,
Gainesville, FL 32611,USA}
\author{N. S. Dalal}
\affiliation{Department of Chemistry and National High Magnetic
Field Laboratory, Tallahassee, FL 32310, USA}
\author{J. M. North}
\affiliation{Department of Chemistry and National High Magnetic
Field Laboratory, Tallahassee, FL 32310, USA}

\date{\today}

\begin{abstract}
We present detailed angle-dependent single crystal electron
paramagnetic resonance (EPR) data for field rotations in the hard
plane of the $S=10$ single molecule magnet Mn$_{12}$-acetate. A
clear four-fold variation in the resonance positions may be
attributed to an intrinsic fourth order transverse anisotropy
($\hat{O}^4_4$). Meanwhile, a four-fold variation of the EPR
lineshapes confirms a recently proposed model wherein disorder
associated with the acetic acid of crystallization induces a
locally varying quadratic (rhombic) transverse anisotropy
[$\hat{O}^2_2\equiv E(\hat{S}^2_x-\hat{S}^2_y)$]. These findings
explain most aspects of the magnetic quantum tunneling observed in
Mn$_{12}$-acetate.
\end{abstract}


\maketitle

Nanometer scale single molecule magnets (SMMs), consisting of a
core of exchange-coupled transition metal ions ({\em e.g}. Mn, Fe,
Ni or Co, {\em etc}.) that collectively possess a large magnetic
moment per molecule (thus far up to approximately $26\mu_B$), have
attracted considerable recent theoretical and experimental
interest.\cite{Angewandte03,MRS00} SMMs offer a number of
advantages over other types of magnetic nanostructures - for
example, they are monodisperse, thereby enabling fundamental
studies of properties intrinsic to magnetic nanostructures that
have previously been inaccessible. Indeed, recent studies of SMMs
have revealed the quantum nature of the spin-dynamics in a
nanomagnet:\cite{Friedman96,Thomas96,Peer98} a metastable state of
the magnetization, say "spin-up," has been shown to relax by
quantum tunneling through a magnetic anisotropy barrier to a
"spin-down" state, in a process called magnetic quantum tunneling
(MQT). Remarkably, MQT in SMMs can be switched on and off, either
via a small externally applied field\cite{WernsdScience99} or by
chemically controlling local exchange interactions between pairs
(dimers) of SMMs, so-called "exchange-bias".\cite{WernsdNature02}
Hence the promise of future applications involving SMMs, such as
single molecule data storage elements or magnetic
Qbits.\cite{LossNature01}

The essential physics of a SMM originates from its large total
spin {\em S}, and an axial or Ising-like magnetocrystalline
anisotropy.\cite{Angewandte03,MRS00} To a fairly good
approximation,\cite{HillPRL98,HillS903} one can describe such a
giant spin by the following effective spin
Hamiltonian:\cite{Barra97,MirebeauPRL99}

\smallskip

\noindent{\hfill\hfill\hfill\hfill\hfill\hfill\hfill $ \hat H =
D\hat S_z^2 + E\left( {\hat S_x^2  - \hat S_y^2 } \right) + \mu _B
\vec B \cdot
\mathord{\buildrel{\lower3pt\hbox{$\scriptscriptstyle\leftrightarrow$}}
\over g}  \cdot \hat S + \hat O_4  + \hat H'
\hfill\hfill\hfill\hfill\hfill\hfill\hfill (1) \hfill$}

\smallskip

\noindent{where {\em D} $(< 0)$ is the axial anisotropy constant,
and $\hat{S}_z$ is the spin projection operator along the easy-
($z$-) axis of the molecule; the second rhombic term
($\hat{O}^2_2$) characterizes the transverse anisotropy within the
hard magnetic plane of the SMM (here, $\hat{S}_x$ and $\hat{S}_y$
are the {\em x} and {\em y} projections of the total spin operator
$\hat{S}$); the third term represents the Zeeman interaction with
an applied magnetic field $\vec{B}$
($\mathord{\buildrel{\lower3pt\hbox{$\scriptscriptstyle\leftrightarrow$}}
\over g}$ is the Land$\acute{e}$ g-tensor); $\hat{O}_4$ includes
fourth order terms in the effective crystal field ($\hat{O}^2_4$,
$\hat{O}^0_4$ and $\hat{O}^4_4$, {\em
etc}..\cite{Barra97,MirebeauPRL99}); and, $\hat{H}^{\prime}$
denotes environmental couplings such as hyperfine, dipolar and
exchange interactions.\cite{KPark02b,HillPRB02} For the strictly
axial case ($\vec{B}\parallel z$ and $E = \hat{O}_4 =
\hat{H}^{\prime} = 0$), the energy eigenstates may be labeled by
the quantum number M$_S$ $(-S \leq $ M$_S \leq S)$, which
represents the projection of $S$ onto the easy axis of the
molecule. The energy eigenvalues are then given by the expression
$\varepsilon = -|D| $M$_S^2$, resulting in an energy barrier
separating doubly degenerate (M$_S = \pm m$, $m =$ integer $\leq
S$) "up" and "down" spin states. It is this barrier which results
in magnetic bistability and magnetic hysteresis at low
temperatures.\cite{Angewandte03,MRS00} Unlike bulk magnets,
however, this hysteresis, which sets in below a characteristic
temperature called the blocking temperature T$_B$ $(<< D
S^2/k_B)$, is intrinsic to each individual molecule $-$ hence the
term SMM.

For MQT to occur, one requires some interaction in Eq.~(1) which
breaks the axial symmetry, thereby mixing "spin-up" and
"spin-down" states.\cite{Angewandte03,MRS00} Sources of symmetry
breaking include: a finite rhombic $(E-)$ term; transverse nuclear
or dipolar fields (included in $\hat{H}^{\prime}$); or higher
order transverse zero-field interactions such as $\hat{O}^2_4$ and
$\hat{O}^4_4$. While this fundamental requirement for MQT is
obvious, in practice, efforts to identify the sources of symmetry
breaking in many SMMs have been problematic. In this letter, we
present evidence in support of a recently proposed solution to the
long standing problem concerning the origin of the transverse
interactions responsible for MQT in
[Mn$_{12}$O$_{12}$(CH$_3$COO)$_{16}$(H$_2$O)$_4$]$\cdot$2CH$_3$COOH$\cdot4$H$_2$O
(Mn$_{12}$-acetate, or Mn$_{12}$-ac). Mn$_{12}$-ac has been the
most widely studied
SMM,\cite{Angewandte03,MRS00,Friedman96,Thomas96,Peer98,HillPRL98,HillS903,Barra97,MirebeauPRL99,KPark02b,HillPRB02}
and was the first to show quantum steps in its low-temperature
hysteresis loops;\cite{Friedman96,Thomas96} it also has the
highest blocking temperature (T$_B~\sim 2$~K) of any known
SMM.\cite{Angewandte03} These characteristics result from the
large total spin ($S = 10$) of the molecule, and its intrinsically
high symmetry ($S_4$). Indeed, it has long been assumed that the
rhombic term in Eq.~(1) is zero ($\hat{O}^2_4$ also) for
Mn$_{12}$-ac due to the four-fold symmetry of the
molecule.\cite{HillPRL98,Barra97,MirebeauPRL99} For this reason,
many groups have conducted extensive searches for other possible
sources of symmetry
breaking.\cite{ChudnPRL01,KPark02b,HillPRB02,AmigoPRB02,MertesPRL02,delBarcoEPL02,CorniaPRL02}

While high-frequency electron paramagnetic resonance
(EPR)\cite{Barra97} and neutron studies\cite{MirebeauPRL99} have
provided convincing evidence for a fourth-order transverse
zero-field term ($\hat{O}^4_4$), such an interaction cannot
explain many key experimental factors associated with the observed
low-temperature hysteresis loops, {\em e.g}. odd-to-even M$_S$ MQT
steps.\cite{ChudnPRL01} For this reason, recent theoretical and
experimental efforts have focused on the possible role played by
disorder.\cite{ChudnPRL01,KPark02b,HillPRB02,AmigoPRB02,MertesPRL02,delBarcoEPL02,CorniaPRL02}
Indeed, our recent EPR investigations have identified a
significant $D$-strain effect (distribution in {\em D} of up to
$2\%$) in typical Mn$_{12}$-ac crystals,\cite{KPark02b,HillPRB02}
while magnetic relaxation measurements by two independent groups
have provided clear evidence for distributions in the tunnel
splittings in Mn$_{12}$-ac,\cite{MertesPRL02,delBarcoEPL02} {\em
i.e}. there are molecule-to-molecule variations in the magnitudes
of the transverse interactions responsible for MQT. Initial
attempts to explain these observations considered long range
elastic strains, caused by various types of dislocation, thus
producing a broad distribution of tunnel splittings via a weak
spatially-varying rhombic anisotropy.\cite{ChudnPRL01} While the
predicted distributions were considerably broader than those
observed experimentally,\cite{delBarcoEPL02} this theory
demonstrated that local disorder-induced rotations of the SMM
magnetic axes could unfreeze the odd MQT resonances (odd-to-even
M$_S$) that are always seen experimentally. Very recently, on the
basis of detailed X-ray studies, Cornia {\em et al.} have proposed
a more realistic model involving a discrete disorder associated
with the acetic acid of crystallization.\cite{CorniaPRL02} This
disorder also gives rise to a locally varying rhombicity
($E$-strain) and, hence, to a distribution in tunnel splittings.
However, the predicted distribution is much narrower in this
case.\cite{delBarcoEPL02}

The present investigation has been motivated by the discrete
nature of the disorder suggested by Cornia {\em et
al}.\cite{CorniaPRL02} This model considers just a few variants of
the basic Mn$_{12}$-ac unit $-$ six in total, only four of which
have symmetry lower than $S_4$. The situation is further
simplified in that the hard and medium axes of three of the four
rhombically distorted variants should be aligned along either of
two mutually orthogonal directions within the hard plane. These
three "rogue" variants with lower than $S_4$ symmetry ($n=1$,
$n=2$ trans, and $n=3$ in Cornia's model)\cite{CorniaPRL02}
account for $62.5\%$ of the molecules, while the $E$-value for the
fourth rogue variant ($n=2$ cis, $25\%$ of the molecules) is
predicted to be weak in comparison to the other three. Thus, EPR
measurements as a function of field orientation within the hard
plane could, in principle, detect the "$E$-strain" caused by these
three rogue variants of Mn$_{12}$-ac.

High frequency single crystal EPR measurements were carried out
using a millimeter-wave vector network analyzer (MVNA) and a high
sensitivity cavity perturbation technique; this instrumentation is
described in detail elsewhere.\cite{MolaRSI} In order to enable
in-situ rotation of the sample relative to the applied magnetic
field, we employed a split-pair magnet with a 6.2~T horizontal
field and a vertical access. Smooth rotation of the entire rigid
microwave probe, relative to the fixed field, was achieved via a
room temperature stepper motor (with $0.1^\circ$ resolution). The
Mn$_{12}$-ac samples were synthesized using standard
methods.\cite{Lis80} For the angle dependent measurements, a
single needle shaped crystal ($\sim 1.4 \times 0.2 \times
0.2$~mm$^3$) was aligned with the axis of a cylindrical TE011
(49.86~GHz) cavity, such that its magnetic easy axis (needle axis)
approximately coincided with the field rotation axis; small
mis-alignments are easily accounted for during data analysis (see
below). Fixed angle, temperature dependent measurements were
carried out for a second crystal (similar dimensions) in a high
field magnet at the National High Magnetic Field Laboratory
(NHMFL). For this experiment, the sample was placed on the end
plate of an oversized cylindrical cavity, and sample alignment was
only accurate to within $\pm 5^\circ$. In all cases, the
temperature was stabilized relative to a calibrated Cernox
resistance sensor using a combination of heaters and cold helium
gas flow.

The upper panel of Fig.~1 shows two typical EPR spectra obtained
with the field aligned along two different directions within the
hard magnetic plane of the sample. The angle $\phi$ is measured
relative to one of the edges of the approximately square cross
section of the sample (see inset in lower panel). Transmission
minima correspond to EPR absorptions, and have been labeled
according to the scheme originally adopted in
ref.~[\onlinecite{HillPRL98}]: $\alpha m$ resonances correspond to
transitions within tunnel-split M$_S=\pm m$ zero-field doublets,
where the applied field mainly contributes to this tunnel
splitting. From this figure, it is immediately apparent that the
spectra change dramatically with $\phi$. First of all, the peak
positions shift appreciably, as illustrated by means of the
contour plot in the lower panel of Fig.~1, where a pronounced
four-fold pattern is seen in the central positions (in field) of
the main EPR absorptions. This four-fold pattern is intrinsic to
the Mn$_{12}$ molecule, and can be attributed to a quartic
transverse zero-field interaction of the form $(\hat{S}_+^4 +
\hat{S}_-^4)$ [=$\hat{O}_4^4$] in Eq.~(1). Fits to this angle
dependence (solid curves) are consistent with a single value of
the coefficient $B_4^4=3.2\pm0.1\times10^{-5}$~cm$^{-1}$, in
excellent agreement with published neutron
studies.\cite{MirebeauPRL99} These fits also include a small
two-fold correction which is due to unavoidable mis-alignments of
the sample's ($\sim 1^\circ$) easy axis with the field rotation
axis, {\em i.e.} this effect is not related to any intrinsic
transverse anisotropy. We note that the $B_4^4$ value obtained
from this analysis is completely independent of all other
parameters in Eq.~(1) (with the exception of $S$) and, therefore,
represents the most direct confirmation for the existence of this
contribution to the transverse anisotropy in Mn$_{12}$-ac.

Also apparent in the upper panel of Fig.~1 are shoulders on the
high-field sides of the $\phi = -15^\circ$ peaks, which are absent
in the $30^\circ$ data. This splitting of the peaks for the hard
plane spectra has been seen in several samples, and also by other
groups,\cite{AmigoPRB02,CorniaPRL02} but has never been studied
systematically. The $\phi$-dependence of this splitting is shown
in Fig.~2 for the first four EPR peaks in Fig.~1
($\alpha6-\alpha9$): Fig.~2a plots the full-widths at half maximum
(FWHM), deduced graphically, for each of the resonances; Fig.~2b
plots the splitting of the individual peaks obtained by fitting
two Gaussians to each resonance. The data in Fig.~2b are limited
in angle range to between $\phi=30^\circ$ and $165^\circ$ because
the slight sample mis-alignment causes the shoulder to move
outside of the available field window for the smallest $\phi$
values (see Fig.~1).

Both sets of curves in Fig.~2 exhibit the same four-fold angle
dependence. However, the origin of this effect is related to a
disorder-induced two-fold rhombic distortion, as we now explain.
For any individual molecule, the quadratic term in Eq.~(1)
[$E(\hat{S}_x^2 - \hat{S}_y^2)$] produces a two-fold pattern for
field rotations in the hard plane. Thus, an intrinsic quadratic
anisotropy of this form ({\em i.e.} same for all molecules) would
produce a two-fold $\phi$-dependent pattern in the hard axis EPR
peak positions, contrary to the observed four-fold shifts
(Fig.~1). However, in the case of Cornia's model (involving
solvent disorder)\cite{CorniaPRL02} one must average over the
disorder. As discussed above, one expects the $n=1$, $n=2$ trans,
and $n=3$ "rogue" Mn$_{12}$-ac variants to exhibit the strongest
evidence for the disorder-induced rhombic anisotropy. However,
because each acetic acid of crystallization can occupy one of four
equatorial positions equally spaced ($90^\circ$ apart) about the
easy-axis of the molecule, each of these variants will be further
sub-divided into four geometric isomers, related through
$90^\circ$ rotations. The result is a 50/50 mixture of molecules
having their hard axes oriented along one of two orthogonal
directions within the hard plane. Since a $90^\circ$ rotation is
equivalent to a change in the sign of {\em E}, the effect of the
disorder will be to produce a splitting, or broadening of the EPR
peaks ({\em not} an overall shift in the central moment), {\em
i.e.} opposite signs of {\em E} produce opposing shifts in the EPR
peak positions. Furthermore, the $\phi$-dependence should now
exhibit a four-fold behavior, since the splitting/broadening
depends on the difference between two $90^\circ$ phase shifted
two-fold patterns. In order to observe a two-fold behavior, one
would have to be able to selectively probe a sub-set of the
molecules with aligned hard axes. We note that such studies would
be possible below the blocking temperature.\cite{delBarco03}

From the angle dependence in Fig.~2, we can deduce that the hard
axes for the $n=1$, $n=2$ trans, and $n=3$ rogue Mn$_{12}$-ac
variants are oriented somewhere between $50^\circ-60^\circ$, and
between $140^\circ-150^\circ$, relative to one of the square edges
of the sample; the hard axes correspond to directions of maximum
linewidth/splitting. Assuming that the $\phi=0^\circ$ and
$90^\circ$ directions correspond to the principal crystallographic
{\em a} and {\em b} axes, then our findings appear to be in
excellent agreement with Cornia's model (which predicts $60^\circ$
relative to {\em a} and {\em b}).\cite{CorniaPRL02} These values
are also in excellent agreement with recent magnetic
measurements.\cite{delBarco03} From the four-fold pattern in
Fig.~1, we see that the four-fold axes (hard and medium) are
approximately locked to the principal {\em a} and {\em b}
directions, {\em i.e.} the maxima and minima occur along
$\phi=0^\circ$, $90^\circ$, {\em etc.}. Thus, the magnetic axes
associated with the quadratic and quartic interactions are not
aligned in a simple way. However, this is not necessarily to be
expected, since the $\hat{O}^4_4$ interaction is dominated by the
symmetry of the Mn$_{12}$O$_{12}$ molecule, whereas the
$\hat{O}^2_2$ interaction is influenced by the peripheral solvent
molecules.

In addition to $E$-strain, Cornia's model predicts slight
variations in $D$ for each of the rogue Mn$_{12}$-ac
variants,\cite{CorniaPRL02} thereby suggesting one possible
explanation for the $D$-strain which we have previously
reported.\cite{KPark02b,HillPRB02} However, this $D$-strain may
also account for the asymmetry seen in the EPR peaks in Fig.~1.
Larger $D$-values will tend to shift pairs of $E$-strain
split-peaks to higher fields. Thus, it is conceivable that the
central portion of the EPR peaks is comprised of several
overlapping split peaks, whereas the shoulder corresponds to the
upper split-peak of one of the rogue variants which is just
resolved from the main peak. Based on the variations in linewidth
and the maximum splitting, we can place an upper bound on the
$E$-values for the rogue molecules of $E_{max}=0.01$~cm$^{-1}$.
This value is about a factor of three greater than the maximum
value predicted by Cornia {\em et al.},\cite{CorniaPRL02} and is
also in excellent agreement with magnetic relaxation
measurements.\cite{delBarcoEPL02,delBarco03}

Finally, in Fig.~3, we show high-field temperature dependent data
for a second sample. The field was oriented along
$\phi=45^\circ\pm 5^\circ$ for these studies, {\em i.e.} close to
the hard and medium axes for the $n=1$, $n=2$ trans, and $n=3$
rogue Mn$_{12}$-ac variants. Derivatives are plotted so as to
accentuate the splitting of the $\alpha 10$ resonance (not seen in
Fig.~1), which corresponds to the transition from the ground state
of the molecules. Upon cooling, the splitting of the $\alpha 10$
resonance persists, while all of the other peaks vanish; for the
two lowest temperatures, the $\alpha 10$ split-peaks perfectly
overlap. This proves, beyond doubt, that the lineshape is the
result of distinct Mn$_{12}$-ac variants with different $D$ and
$E$ values. We note that the $\beta 10$ resonance has previously
been attributed to disorder in a separate EPR study by Amigo {\em
et al.}\cite{AmigoPRB02} However, the data in Fig.~3 strongly
suggest that $\beta10$ corresponds to an excited state of the
molecule, since it vanishes as T$\rightarrow 0$. The origin of
these $\beta$ transitions (which we have previously
reported\cite{HillPRL98}) will be discussed in a separate
publication.\cite{HillS903}

In summary, we present single crystal EPR data for field rotations
within the hard-plane of Mn$_{12}$-ac. A clear four-fold variation
in the resonance positions can be attributed to an intrinsic
fourth order transverse anisotropy ($\hat{O}^4_4$). Meanwhile, a
four-fold variation of the EPR lineshapes confirms a recently
proposed model\cite{CorniaPRL02} wherein solvent disorder induces
a locally varying quadratic anisotropy ($\hat{O}^2_4$). We note
that very recent magnetic relaxation experiments clearly indicate
that it is probably this $E$-strain which dominates the magnetic
quantum tunneling in Mn$_{12}$-ac.\cite{delBarco03}

We thank A. Kent, E. delBarco, G. Christou and D. Hendrickson for
useful discussions. This work was supported by the NSF
(DMR0103290, DMR0196430 and DMR0239481); the NHMFL is supported by
the State of Florida and the NSF under DMR0084173. S. H. would
like to thank the Research Corporation for financial support.

\clearpage

\noindent{{\bf Figure captions}}

\bigskip

FIG. 1. Upper panel: typical EPR spectra obtained at 49.86~GHz and
T$= 15$~K, with the field aligned along two different directions
within the hard magnetic plane of the sample (indicated in the
figure); the angle $\phi$ is measured relative to one of the
square edges of the sample. The resonances have been labeled
according to a previously developed scheme.\cite{HillPRL98} Lower
panel: color contour plot illustrating the angular variation of
the peak positions in the upper panel; data were taken at
$15^\circ$ intervals (red indicates strong absorption). The dashed
line corresponds to the blue curve in the upper panel, while the
$y$-axis corresponds to the red curve. The inset depicts the
geometry of the experiment.

\bigskip

FIG. 2. Hard plane angular variation of (a) the linewidths, and
(b) the line splitting, for the four strongest EPR peaks in Fig.~1
($\alpha 6$ to $\alpha 9 $).

\bigskip

FIG. 3. Temperature dependence of the derivatives of high field
spectra with the field applied at $\phi = 45^\circ \pm 5^\circ$
within the hard plane; the temperatures are indicated in the
figure.

\bigskip

\clearpage
\begin{figure}

\includegraphics{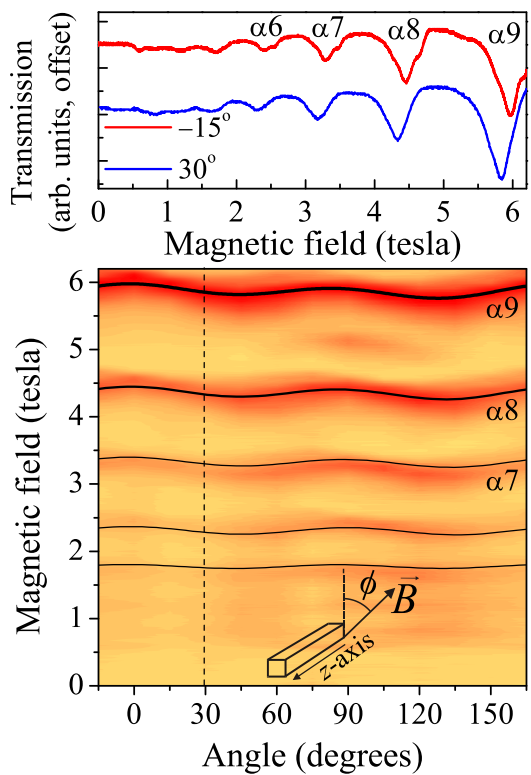}
\caption{\label{fig1} S. Hill {\em et al.}}
\end{figure}

\bigskip

\begin{figure}

\includegraphics{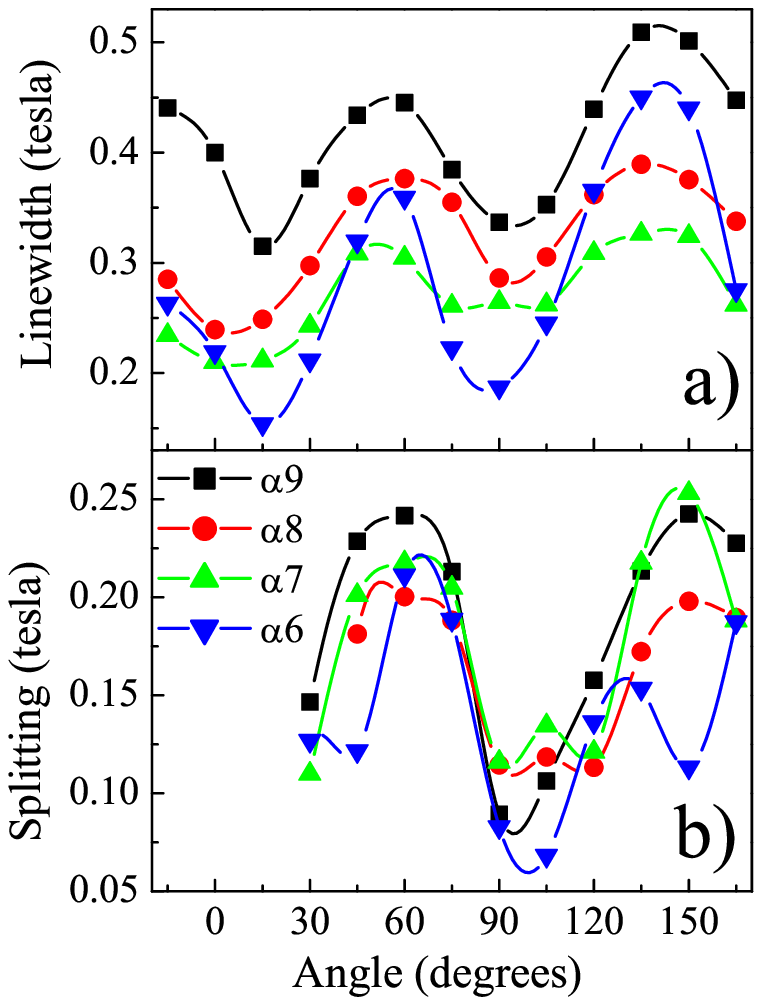}
\caption{\label{fig2} S. Hill {\em et al.}}
\end{figure}

\bigskip

\begin{figure}

\includegraphics{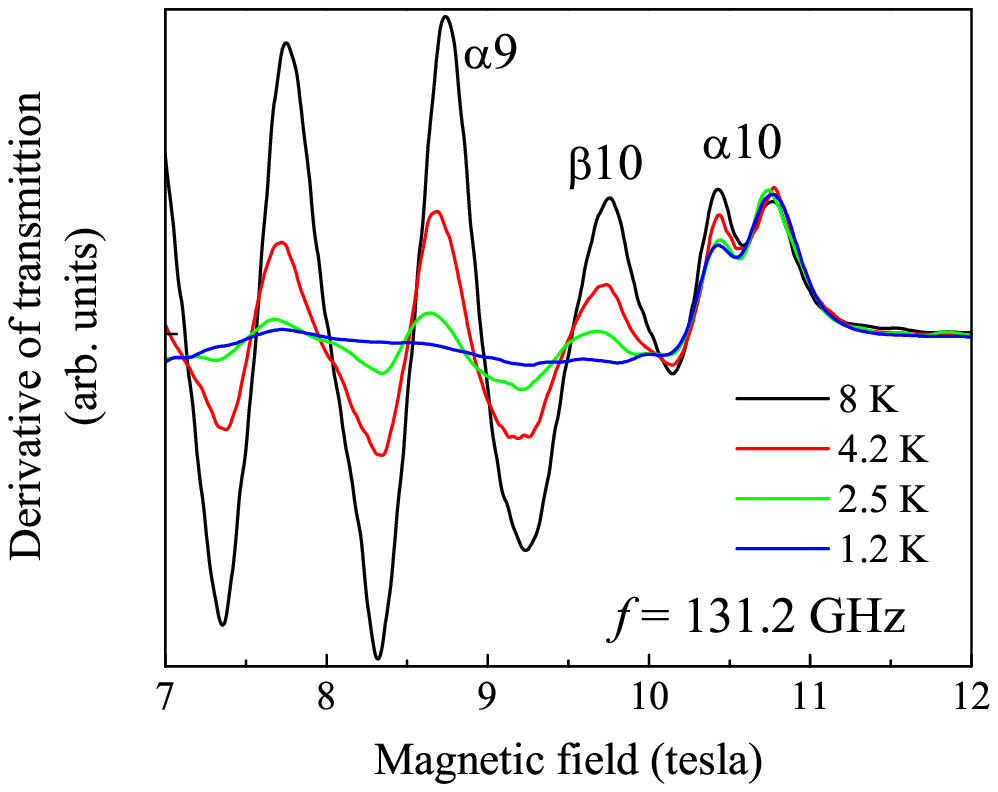}
\caption{\label{fig3} S. Hill {\em et al.}}
\end{figure}


\begin{thebibliography}{10}

\bibitem{Angewandte03}
D. Gatteschi and R. Sessoli, Angew. Chem. {\bf 42},  268  (2003).

\bibitem{MRS00}
G. Christou {\it et~al.}, MRS Bulletin {\bf 25},  66  (2000).

\bibitem{Friedman96}
J.~R. Friedman {\it et~al.}, Phys. Rev. Lett. {\bf 76},  3830
(1996).

\bibitem{Thomas96}
L. Thomas {\it et~al.}, Nature {\bf 383},  145  (1996).

\bibitem{Peer98}
J.~A. A.~J. Perenboom {\it et~al.}, Phys. Rev. B {\bf 58},  330
(1998).

\bibitem{WernsdScience99}
W. Wernsdorfer and R. Sessoli, Science {\bf 284},  133  (1999).

\bibitem{WernsdNature02}
W. Wernsdorfer {\it et~al.}, Nature {\bf 416},  406  (2002).

\bibitem{LossNature01}
M.~N. Leuenberger and D. Loss, Nature {\bf 410},  789  (2001).

\bibitem{HillS903}
R.~S. Edwards {\it et~al.}, cond-mat/0302052 (unpublished).

\bibitem{HillPRL98}
S. Hill {\it et~al.}, Phys. Rev. Lett. {\bf 80},  2453  (1998).

\bibitem{Barra97}
A.~L. Barra, D. Gatteschi, and R. Sessoli, Phys. Rev. B {\bf 56},
8192
  (1997).

\bibitem{MirebeauPRL99}
I. Mirebeau {\it et~al.}, Phys. Rev. Lett. {\bf 83},  628  (1999).

\bibitem{KPark02b}
K. Park {\it et~al.}, Phys. Rev. B {\bf 66},  144409  (2002), also
Phys. Rev. B
  {\bf 65}, 14426 (2002).

\bibitem{HillPRB02}
S. Hill {\it et~al.}, Phys. Rev. B {\bf 65},  224410  (2002).

\bibitem{ChudnPRL01}
E.~M. Chudnovsky and D.~A. Garanin, Phys. Rev. Lett. {\bf 87},
187203  (2001),
  also Phys. Rev. B {\bf 65}, 094423 (2002).

\bibitem{AmigoPRB02}
R. Amig{\'o} {\it et~al.}, Phys. Rev. B {\bf 65},  172403  (2002).

\bibitem{MertesPRL02}
K.~M. Mertes {\it et~al.}, Phys. Rev. Lett. {\bf 87},  227205
(2001).

\bibitem{delBarcoEPL02}
E. del Barco {\it et~al.}, Europhys. Lett. {\bf 60},  768  (2002).

\bibitem{CorniaPRL02}
A. Cornia {\it et~al.}, Phys. Rev. Lett. {\bf 89},  257201
(2002).

\bibitem{MolaRSI}
M. Mola {\it et~al.}, Rev. Sci. Inst. {\bf 71},  186  (2000).

\bibitem{Lis80}
T. Lis, Acta Cryst. B {\bf 36},  2042  (1990).

\bibitem{delBarco03}
E. del Barco {\it et~al.} arXiv:cond-mat/0304206.

\end{thebibliography}
\end{document}